# Behind the Eastern-Western European convergence path: the role of geography and trade liberalization

**Adolfo Cristobal-Campoamor**[1]

**Osiris Jorge Parcero**[2]

**Keywords**: Trade liberalization; transition; convergence; Eastern Europe

**Abstract**

This paper proposes a 2-block 3-region economic geography model that can account for the most salient stylized facts experienced by Eastern European transition economies during the period 1990-2005. In contrast to the existing literature, which has favored technological explanations, trade liberalization is the only driving force. The model correctly predicts that in the first half of the period trade liberalization led to divergence in GDP per capita, both between the West and the East and within the East. Consistent with the data, in the second half of the period, this process was reversed and convergence became the dominant force.

[1] A. Cristobal-Campoamor, Graduate School of Economics and Management, Ural Federal University, Yekaterinburg, Russia, e-mail: adolfocristobal@gmail.com

[2] O. J. Parcero, Department of Economics and Finance, College of Business and Economics, United Arab Emirates University, Al Ain, United Arab Emirates, e-mail: osirisjorge.parcero@gmail.com

# 1. Introduction

From the beginning of the 1990s to 2005 the Eastern European economic transition has been characterized by a U-shaped pattern of relative development (see Figure 1). Initially, relative income per capita between Eastern and Western Europe diverged, but roughly from 1999 onwards this pattern was reversed and Eastern Europe started to catch up with its Western counterpart. Moreover, when analyzing the performance inside Eastern Europe a similar pattern emerges. The countries closer to the West (Border) initially experienced faster growth than those situated further from the West (Hinterland), but from the end of the 1990s this was also reversed.

A similar pattern can be observed when the attention is focused on industrial output. Hinterland (East) initially suffered a continuous relative deindustrialization, followed by a remarkable recovery from the beginning of the new century (see Figures 2 and 3). Thus, the stylized facts exhibit both an East-West and a Border-Hinterland U-shaped pattern in terms of (real) GDP per capita and industrial output.

The literature has typically explained these U-shaped patterns by relying on technological arguments or on the misallocation of factors of production. Boldrin and Canova (2003), for example, suggest that the technological obsolescence led to an initial period of intense unemployment and reallocations after trade was liberalized. Blanchard (1996) and Blanchard and Kremer (1997) link the initial slump to microeconomic "disorganization": the collapse of the state sector was precipitated by traditional input suppliers, who found attractive opportunities outside the state sector and broke the established productive chains. Cociuba (2006) and Keller (1997) also stress the role played by technology adoption to account for the GDP trajectories of Eastern European countries. The existing literature thus puts the emphasis on the intensity of reallocations that were needed to adapt to a superior Western technology, followed by a remarkable catch-up process that was conditioned by redistributive public policies.

While we do not claim these explanations are erroneous in any way, in this paper we deliberately disregard issues of technological backwardness or sectorial misallocations. Instead, we propose an economic geography model where trade liberalization is the only driving force. This is based on our belief that the trade reorientation towards the West and the initial deterioration of the market for exports were crucial initial conditions for the transition (see, among others, Campos and Coricelli (2002), Christoffersen and Doyle (1998), de Melo et al. (2001) or Tondl and Vuksic (2003)). Our focus is on the evolution of the Eastern-European economies right after the collapse of the socialist experiment. Besides, we hope the insights derived from our experiment could be relevant to analyze other trade liberalization processes in different countries. Since it can be argued that, prior to the transition period, all Central and Eastern European countries shared similar technological and institutional conditions, we claim that the disparate evolutions of Border and Hinterland may have something to do with the relative proximity to the EU market. That is the reason why we emphasize the causal determinants related to geography and trade openness.

We develop a model with 2 blocks (West and East) and three regions (one region in the West, and a Border and Hinterland in the East). As is usually the case in the Economic Geography literature, our model assumes that agriculture is perfectly competitive, industry is monopolistically competitive and workers are perfectly mobile between sectors. Trade in industrial goods is subject to transport costs, which are higher between Hinterland and West than between Border and West. As in Krugman and Venables (1995) and Puga (1999), industrial firms use intermediate goods, which gives rise to forward and backward linkages. The relative size of the endowments of labor and land in our model match the actual shares in our three regions. For simplicity and to make our results as sharp as possible, we assume that West and East exhibit identical technologies.

This simple setup, which abstracts from technological differences, is sufficient to account for the main stylized facts. The focus of our analysis is the pattern of convergence in terms of GDP and industrialization (between East and West; Hinterland and Border) produced by the trade liberalization during the transition period. Our model can be viewed as a generalization of Krugman and Venables (1995) and Puga (1999). They showed how in a two- block model the earlier stages of trade liberalization could bring about lower real wages and deindustrialization in disfavored markets.

Our paper differentiates from Puga (1999) in that we allow for internal trade costs within the East, making the setup non-equidistant. In that way Border and Hinterland become asymmetric in terms of their distance to the West. For that reason, we must rely on numerical simulations. This modification is not substantial from a purely theoretical point of view and so our paper should be considered only an empirical application of Puga (1999), with a slight structural modification. A similar exercise was undertaken by Bosker et al. (2010), with the qualification that they did not focus on the historical experience of transition, but they tried to shed light on the future impact (and only within the West) of the recent EU enlargement.

Our paper is a contribution to this literature.

Our simulation results explain very well the actual relative evolution of the different regions described above. First, during the time following the trade liberalization these results are as follows. Trade liberalization should initially lead to divergence in GDP per capita, both between West and East and between Border and Hinterland. The good performance of the West can be explained by a Home Market effect. As trade costs drop, the relative profitability varies in favor of the largest market. This same phenomenon leads to the relative initial deindustrialization of the East in favor of the West. Furthermore, the results show that there should be an initial divergence between Border and Hinterland in favor of the former location. This is the case because the proximity to the West involves a crucial access to the bulk of consumption goods and intermediate inputs.

Second, the results show that during the final stages of East-West trade liberalization a convergence in GDP per capita should appear, both between Border and Hinterland and between East and West. On the one hand, this is the case because when international trade costs are sufficiently low, proximity to large markets is no longer a basic determinant of firms' location. On the other hand, the previous agglomeration in the West led to (relatively) lower wages in the Border and even more so in the Hinterland. This is now crucial for the recovery of a substantial manufacturing basis by the Border and to a larger extent by the Hinterland.

We also present a series of industry-level stylized facts to give additional support to our hypothesis that NEG forces were at work during the period. When we break down our output data by industry we observe that, as trade freeness increases, sectors with significant scale economies and/or high value-to-weight ratios follow an inverted U-shaped pattern in their geographical output concentration (first three columns of Figure 13). The spatial concentration patterns of these sectors are strongly affected by economic geography forces. In contrast, sectors where scale economies are less relevant and value-to-weight ratios are lower (last three columns of Figure 13) follow a monotonic concentration pattern. The latter is more consistent with theories of comparative advantage. We argue that these stylized facts seem to be more consistent with a NEG approach and less so with the conventional technological-obsolescence plus catch-up approach.

The paper is organized as follows. Section 2 mentions some additional related literature. Section 3 defines the three regions and further describes the policy changes and stylized facts, justifying the adoption of the assumptions underlying our policy experiment. Section 4 briefly describes the analytical framework, which is more extensively presented in the Appendix. Section 5 uses numerical experiments to study the effects of trade liberalization. Section 6 shows some microeconomic stylized facts. Section 7 concludes.

## 2. Other related literature

In addition to the papers cited in the introduction the following literature is worth mentioning. Some other papers have already developed more-than-two-region models of economic geography, either using the Dixit and Stiglitz's modeling framework (Krugman and Livas-Elizondo (1996), Alonso-Villar (1999), Monfort and Nicolini (2000), Crozet and Koenig-Soubeyran (2004) etc.) or using the Otaviano-Tabuchi-Thisse's one (Behrens et al. (2006)). Ago et al. (2006) uses both methods and compares them in the context of a three-region model.

Another important paper is Venables (2000), who uses a three-location framework as well, but focuses on the internal geography of a developing country that is hardly industrialized for intermediate levels of trade costs. Venables' insights are probably more applicable to countries like China and India, which - given their lower initial level of development - could not possibly experience divergence with respect to the West during their transition process. Our starting point is different because for all levels of trade costs, both West and East are significantly industrialized.

Nevertheless, to the best of our knowledge, only Forslid et al. (2002), Brülhart et al. (2004), Crozet and Koenig-Soubeyran (2004) and Iranzo and Peri (2009) were directly motivated by the experience of transition economies. Both Brulhart et al. (2004) and Forslid et al. (2002) present scenarios for the *future* economic geography of Europe, but without a retrospective approximation to the 1990s. In particular, Brulhart et al. (2004) pay attention to the internal geography of the West as a result of the EU enlargement. In contrast to their paper, our priority is to explore the patterns of industrial relocation within the East, together with their implications for relative (real) GDP per capita, during the period 1990-2005.

In the case of Crozet and Koenig-Soubeyran (2004), the authors also focus on the connections between external trade and internal geography, showing that both the concentration and the dispersion of industry are theoretically possible for Eastern Europe. However, they completely disregard convergence issues. These authors link the mass of manufacturing varieties to the (fixed) national stock of mobile labor (as in Krugman (1991)). For this reason they cannot study the processes of industrialization/deindustrialization at the national level, and cannot reproduce our non-monotonic patterns of convergence.

Finally, Iranzo and Peri (2009) examine the welfare consequences of the liberalization of both trade and East-West migration, with the first reform predating the second. For them, trade liberalization is a one-shot event instead of a gradual process. This implies that the non-monotonic patterns of convergence generated by trade openness cannot be captured by their model. Our model tries to fill this gap.

## 3. Introducing the scenario under analysis

*3.1. Definition of the three regions*

The different U-shapes mentioned in the introduction are robust to different criteria in the selection of Border and Hinterland countries. For instance, we could place in the Border those countries sharing boundaries with an EU-15 thriving economy. Alternatively, we could just call Border to the set of countries that joined (or will join soon, like Croatia) the European Union.[3] Therefore, and without big loss of generality, we decided to choose the latter convention and specify our regions as follows: West (EU-15), Border (Bulgaria, Croatia, Czech Republic, Estonia, Hungary, Latvia, Lithuania, Poland, Romania, Slovak Republic and Slovenia) and Hinterland (Albania, Armenia, Azerbaijan, Belarus, Georgia, Kazakhstan, Kyrgyz Republic, Macedonia, Moldova, Russian Federation, Tajikistan, Turkmenistan, Ukraine and Uzbekistan).[4]

Our choice of the national (instead of the sub-national) level as the relevant spatial dimension to characterize our stylized facts has been deliberate. In that respect, we are consistent with the empirical literature, which emphasizes how in Eastern Europe country-specific factors are more important determinants of regional income differences than localized conditions (see Brulhart and Koenig-Soubeyran (2006),

---

[3]The simulations run under this alternative specification are available upon request.
[4]Bosnia, Montenegro and Serbia are not considered because not all data were available since 1990.

Melchior (2008) or Bosker (2009)).

*3.2. Policy changes*

*West-Border trade liberalization*

The route towards East-West trade liberalization started quite early in some Eastern-European countries like the former Yugoslavia and Romania. In particular, the European Community signed an initial Generalized System of Preferences with Romania in 1974, and an agreement on manufacturing trade was reached in 1980. Nevertheless, the most comprehensive Generalized Systems of Preferences (GSP) were approved by the EU and individual Eastern countries at the beginning of the 1990s. The EU granted GSP status first to Hungary and Poland (1990), then to Bulgaria and former Czechoslovakia (1991), and subsequently to Estonia, Latvia and Lithuania (1992). All these reforms culminated with the accession of all the Border countries to the European Union (Croatia is expected to do it very soon).

In Figure 4 we display the evolution of the three region-pairwise macroeconomic trade freeness (using data from IMF Bilateral Trade Statistics). Moreover, in Figure 5 we also display the West-Border evolution of the medians of the trade freeness across industrial sectors (using data from Nicita and Olarreaga (2007)). The concept of trade freeness, used by Head and Mayer (2004), is an inverse measure of the magnitude of the trade costs between two particular locations. It is computed in the following way:

$$\emptyset_{ij} = \left(\frac{m_{ij} m_{ji}}{m_{ii} m_{jj}}\right)^{1/2}$$

where $\emptyset_{ij}$ denotes the trade freeness between locations $i$ and $j$; $m_{ij}$ stands for country $i$'s imports of country $j$'s manufacturing goods; and $m_{ii}$ represents country $i$'s shipments to itself, computed as the difference between the value of the gross output in the producing country minus the aggregate value of its exports.

Figures 4 and 5 show how remarkable was the rise of the aggregate and industry-level West-Border's trade freeness during the period under consideration. On the contrary, the aggregate trade freeness remained quite stable between West-Hinterland and Border-Hinterland. Unfortunately, the inexistence of industry-level data on gross output and bilateral trade for the Border and Hinterland countries prevented us from calculating the evolution of the industry-level West-Hinterland and Border-Hinterland's trade freeness.

*Limited migration*

According to Kaczmarczyk and Okolski (2005), during the communist era migration in Eastern Europe was negligible, both within and between countries. Rural-to-urban mobility was also greatly delayed and generally low. Moreover, in contrast to Western European nations, in many Eastern countries the process of industrialization took place in the absence of massive urbanization.

It was during the 1990s that substantial policy reforms were enacted to liberalize labor flows across Eastern European countries. For example, in 1993 the Czech Republic established a liberal migration policy which turned the country into the home to tens of thousands of migrants from Europe and Asia (Drbohlav (2005)). In 1993 Russia abolished the internal passport and allowed for freedom of movement (Heleniak (2002)). This also resulted in many migrants coming from other Hinterland countries.

However, the magnitude of the Hinterland-Border permanent migratory movements was relatively insignificant. According to Mansoor and Quillin (2006) "there are minimal flows from the CIS [Hinterland] states into the [Border]", which amount approximately to 5% of the CIS countries total emigration flow. Such flow from the Hinterland to the Border could be quantified for the year 2003 in about 300 thousand people, out of 6 million people who permanently emigrated from their Hinterland country in that particular year.

The migration from Border to Western European counties between 1990 and 2000 faced the existence of many restrictions, and the migration from Hinterland to West was almost inexistent. In this context a notable exception was the migration of Eastern European ethnic Germans towards their homeland. Hence, the stock of Border's immigrants in Western Europe only rose from 758,193 to 965,724 in the period 1990-2000 (see Pytlikova (2006)). However, from the turn of the century the Border-West migration gained significance. According to Boeri and Brücker (2005), the stock of immigrants increased by 1.1 million between 1990 and 2005. Thus, in the five-year period 2000-2005 the stock of immigrants has roughly increased 4.3 times more than it did in the previous decade, though the 1.1 million represents only 1% of Border's population in 1990.

*3.3. Main stylized facts*

The three stylized facts we aim to account for in our theoretical model are:

i. *U-shaped pattern of relative income per capita between East and West:* in Figure 1 we see that the relative income per capita between East and West diverges until 1999 and starts to converge thereafter.

*ii.U-shaped pattern of relative income per capita within East (Hinterland relative to Border):* this can be seen in Figure 1, where the turning point is again around 1999.

*iii.U-shaped pattern of industrialization between East and West and between Hinterland and Border:* Figure 6 shows the industrial output's share of both Border and Hinterland as a fraction of the aggregate European industrial output. It is apparent that the Border keeps a roughly stable share, whereas the Hinterland's share initially decays sharply but firmly recovers later. We recognize that it is a frequent suspicion that the industrial and GDP revival of the Hinterland may be linked to the "petro-boom" experienced by resource-rich countries like Russia. However, we show in Figures 7 and 8 that the U-shaped patterns are robust to the exclusion of the Russian Federation.

## 4. The model

In order to make the reading easier and given that our model is an application of Puga (1999), we have decided to move its lengthy mathematical description to the Appendix. This model considers a framework with two blocks (West and East) and three regions: West ($W$), and Border ($B$) and Hinterland ($H$) in East. There are two sectors, agriculture and industry; two factors of production, labor and land; and both blocks have identical technologies. East is a larger block in terms of land area and population, though the Western market is better integrated due to the absence of internal trade costs.

Table 1 shows each region $i$ endowment's shares $L_i$ and $K_i$ (for population and land area respectively) of the three regions' aggregate. The shares for the year 1990 are used, $L_{i90}$ and $K_{i90}$,[5] except for labor in West and Border, which are allowed to change according to the migration flows. For the latter case, it is assumed that individuals get a disutility from emigration and hence they do not emigrate unless such disutility is offset by real-wage differentials (see the Appendix for more details). $\psi$ is the share of the 1990-Border's population that remains in Border (a function of West and Border real wages) and $m$ is a migration openness' indicator taking the value of 1 if Border-West migration is allowed and 0 otherwise.[6]

**Table 1**

|  | **Regional population shares** | **Regional land area shares** |
|---|---|---|
| **West** | $L_W = \psi_W L_{W90} + (1-\psi_B)L_{B90}$   $if\ m = 1$ <br><br> $L_W = L_{W90} = 0.48$   $if\ m = 0$ | $K_{W90} = 0.12$ |
| **Border** | $L_B = \psi_B L_{B90} + (1-\psi_W)L_{W90}$   $if\ m = 1$ <br><br> $L_B = L_{B90} = 0.15$   $if\ m = 0$ | $K_{B90} = 0.04$ |
| **Hinterland** | $L_{H90} = 0.37$ | $K_{H90} = 0.84$ |

---

[5] These shares are obtained from the World Development Indicators and remained very stable along the entire time period. Moreover, notice that with the exception of $L_{i90}$ and $K_{i90}$, we omit a subscript $t$ to refer to time. This is done for the sake of simplicity and to be in line with Puga's notation.

[6] For simplicity of exposition, Table 1 assumes that the migration stock West-to-Border is non-negative, i.e. $L_W \geq L_{W90}$ at any time. In fact this is the case in the simulations, even though $L_W < L_{W90}$ is allowed.

Trade in industrial goods is subject to different transport costs for each region pairwise (details in the next section). The agricultural good is produced under perfect competition, using land and labor as inputs. Since the supply of land is fixed, the agricultural sector faces decreasing returns to labor. This entails that agriculture endogenously takes place in all locations. The monopolistically competitive industrial sector uses labor and intermediate goods, giving rise to forward and backward linkages. Our model is very much similar to Puga (1999), though it exhibits two basic differences. That is, the introduction of transport costs between Border and Hinterland, which prevents the full convergence of Eastern and Western welfare levels, even under perfect trade openness between East and West. The second difference is the introduction of migration disutility.

## 1. Numerical simulations

### 1.1. Parameterization

The goal of this section is to carry out an experiment that looks at the effect that a gradual decrease in West-Border trade costs (due to trade liberalization) has on GDP per capita and industrial convergence.

The values for our model´s parameters are mainly taken from Bosker et al (2010), who empirically estimated the parameters in Puga (1999)'s model. These authors used a sample of 194 EU-15 NUTS-II regions over the period 1992-2000 and then applied such estimation to simulate the impact (on the West) of the ongoing integration with the Border countries. As it was stated earlier, our aim is to replicate the stylized facts (1), (2) and (3) by focusing on the pure effects of geography and trade liberalization without resorting to taste or technological differences between the East and the West. Therefore, the structural-parameter-estimates, described in

the Appendix, are as follows:

| | | |
|---|---|---|
| $\sigma =$ | Elasticity of substitution between manufacturing varieties | $= 7.122$ |
| $\gamma =$ | Share of consumers' income spent on manufactures | $= 0.7$[7] |
| $\mu =$ | Share of firms' revenue spent on intermediates | $= 0.284$ |
| $\theta =$ | Labor elasticity in the agricultural production function | $= 0.234$ |

Trade in industrial goods is subject to transport costs, represented by the parameters $\emptyset_{W-B}$, $\emptyset_{B-H}$ and $\emptyset_{W-H}$ with $\emptyset_{W-H} = \emptyset_{W-B} \times \emptyset_{B-H}$ and where $\emptyset = 0$ ($\emptyset = 1$) can be translated as to an infinite (zero) trade cost. For our simulations we assume that $\emptyset_{B-H}$ remains constant throughout the whole period at 0.01 while $\emptyset_{W-B}$ gradually increases from 0 to 1.

In our simulations we do not intend to capture the whole set of stable (and unstable) equilibria at any level of the trade costs. We just obtain an initial West-East autarky equilibrium, $\emptyset_{W-B} = \emptyset_{W-H} = 0$, and then, as the trade costs fall, we increase (decrease) the local mass of manufacturing varieties in those locations where profits are positive (negative) until the zero-profit condition is satisfied. Thus, a sequence of equilibria is obtained.

### 1.2. Trade liberalization

---

[7] Bosker et al (2010) use $\gamma = 0.994$, which reflects the share of income spent on anything but agricultural products for the EU-15. At the same time they consider the possibility of using $\gamma = 0.335$, which is the share of industrial goods on aggregate consumption. We have considered more appropriate to report results for $\gamma$ between these two values, but our results are robust to the choice of $\gamma = 0.994$.

The results for our scenario can be seen in Figures 9 and 10, where the horizontal axis shows the (increasing) degree of West-Border trade freeness. Initially Eastern and Western markets are completely isolated from each other. As $\emptyset_{W-B}$ and at a lesser extend $\emptyset_{W-H}$ start to rise, tougher competition appears in both Eastern locations (especially for the Border), whereas the Western market is hardly affected by the scant competitors in the Border and the more distant firms in the Hinterland. This leads to a process of East-West divergence. Nevertheless, openness to the Western market not only implies tougher competition, but also higher exports and cheaper imports of intermediates in the East, especially in the Border. The proximity to a large market is crucial for the Border to absorb a portion of Hinterland's manufacturing share, leading to a process of within-East divergence.

Once trade openness between our three locations starts to be sufficiently high, being closer to large markets – with abundant purchasing power– ceases to be the firms' main priority. Simultaneously, the previous Eastern *deindustrialization* creates a reservoir of cheap labor in the primary sector of the East, especially in the Hinterland, whereas the Western labor costs have been growing higher and higher. Lower nominal wages in the East, together with the possibility to import Western intermediates more cheaply, end up precipitating the consecutive reindustrialization of both Eastern locations: starting with the Border and following with the Hinterland, once the wages in the former increase. Remoteness was Hinterland´s initial disadvantage but, once relative wages became lower there and land rents more abundant, industrial profitability returned. As a robustness check, in Figures 11 and 12 we show the simulation results excluding Russia, which show no significant differences.

# 6. Microeconomic evidence at the industry level

This section uses industry-level evidence to give additional support to our hypothesis that NEG forces were at work during the period. The argument is that the industries exhibiting higher scale economies and/or value-to-weight ratios are more likely to face a spatial concentration of their output, hence being more affected by the centripetal and centrifugal forces characteristic of the NEG. Therefore and as suggested by Forslid et al (2002), for intermediate values of the sectorial trade freeness, if such forces were the predominant ones, these sectors should tend to concentrate in the larger market.

Our task faced a large scarcity of data on industry-level output and bilateral trade for the Border, and their inexistence for the Hinterland. However, we have managed to single out a few country-pairwise industry-level output-concentration ratios for some relevant West-Border pairs of countries. We have graphed them against the industry-level values of trade freeness.

Figure 13 shows that sectors with important scale economies and/or high value-to-weight's ratios (first three columns) display an inverted U-shaped pattern in their output concentration in the West as trade freeness increases. In contrast, sectors where scale economies are less relevant and value-to-weight ratios are lower (last three columns) follow a monotonic concentration pattern. The latter is more consistent with theories of comparative advantage. Our conclusions are based on the sectorial scale economies' ranking taken from Pratten (1988) as well as the value-to-weight ratio from the US-Canada Trade Statistics - North American Trans-border Freight data (NATBF).

A technological-obsolescence approach would not predict the above stylized facts,

except in the case that the sectors with higher economies of scale and/or higher value-to-weight ratios were also the ones which have suffered from a higher technological obsolescence. However, it seems difficult to think that, for instance, Leather and Pottery China have suffered a high obsolescence. Similarly, the ex-communist countries were quite advanced in the production of Non-Ferrous Metals and even Chemicals and so they do not seem to represent their relatively more obsolete sectors.

## 7. Conclusion

In this paper we have examined some mechanisms – exclusively related to economic geography and trade openness – as possible ingredients to account for the relative income profile (and industrialization) of transition countries along the period 1990-2005. We have deliberately disregarded any consideration of technological differences or unrelated public-policy factors, hence our explanations can be considered as complementary to those presented in the literature.

Initially, higher trade openness generates a deterioration of the industrial capacity in the East in favor of the West, due to the higher integration of the market in the latter location. Something similar happens between Hinterland and Border, where foreign inputs are much cheaper. As a result, the early stages of trade liberalization are characterized by both East-West and Border-Hinterland divergence. Later, lower marginal costs channel a recovered manufacturing profitability towards the East, first to the Border and later to the Hinterland, which finally gets reindustrialized under our parameterization. This fact also drives the revival of the East relative to the West.

up? The role of foreign investment, human resources and geography". IEF Working Paper, No. 51.

Venables (1996). Equilibrium locations of vertically linked industries. *International Economic Review*, 37( 2) May, pp. 341-359.

Venables A., 2000, "Cities and trade: external trade and internal geography in developing countries", in S. Yusuf, S. Evenett and W. Wu (eds.), Local dynamics in an era of globalization: 21st century catalysts for developments, OUP and World Bank, Washington DC.

# Appendix

Our model is a slight extension of Puga (1999)'s model. $K_i$ and $L_i$ are the stocks of land and labor in each region $i$, normalized as shares of the total stocks in the global economy. Moreover, the labor stocks of Border and West are adjusted by the evolution of migration.

1. Demand

There is a food sector ($F$), the *numeraire*, and an industrial sector. Consumers have Cobb-Douglas preferences over the primary good and a CES composite of manufacturing *varieties* with an industrial expenditure share $\gamma$, $0 < \gamma < 1$. Maximizing the CES sub-utility function subject to the income constraint yields the consumer demand for each variety $h$ (A.1), where Y is the consumer's income, $q$ is the manufactures' price index, and $\sigma$ and $p(h)$ are the varieties' elasticity of substitution and c.i.f. price respectively.

$$c(h) = p(h)^{-\sigma} q^{\sigma-1} \gamma Y; \quad q = \left(\int_0^n p(h)^{1-\sigma} dh\right)^{1/(1-\sigma)} \quad (A.1)$$

Firms also use varieties as intermediates for which they also have an elasticity of substitution σ. The local demand for a variety depends on the spending by both local consumers and local firms. The spending from location $i$ on any variety is $e_i = \gamma Y_i + \mu n_i p_i x_i$, where $n_i$ ($x_i$) is the mass of produced varieties (the equilibrium output of any variety) in region $i$. The first term is the share of local consumers' income spent on manufactures and the second is the expenditure on intermediates.

### 2. Manufacturing supply

The manufacturing sector is monopolistically competitive. The production input in manufacturing is a Cobb-Douglas composite of labor (with a wage rate of $w_i$) and intermediates. Total cost in region $i$ is:

$$C(x_i) = (q_i^\mu w_i^{1-\mu})(\alpha + \beta x_i) \tag{A.2}$$

where $\alpha$ and $\beta$ are the fixed-cost and marginal-input requirements for each variety. Profit maximization results in a constant mark-up over marginal costs, as shown in (A.3): [8]

$$p_i = (q_i^\mu w_i^{1-\mu}) \tag{A.3}$$

Finally, using the zero-profit condition and the mark-up pricing rule (A.3) we get that the break-even supply of any variety is $x_i = 1$.

### 3. Supply of food

Consumers' income comes from two sources, local workers' wages and agricultural rents. Agriculture produces a homogeneous good under constant returns to scale, perfect competition and no transport cost across locations. Its production in region

---

[8] We have applied the normalizations $\alpha = \frac{1}{\sigma}$ and $\beta = \frac{\sigma-1}{\sigma}$.

$i$ depends on the available local land ($K_i$) and labor ($L_{F,i}$) such that

$$F = F(L_{F,i}) = (L_{F,i})^\theta K_i^{1-\theta} \tag{A.4}$$

By applying rent maximization, the aggregate agricultural rents in location $i$ are

$R_i = (1-\theta)K_i \left(\frac{\theta}{w_i}\right)^{\frac{\theta}{1-\theta}}$, which added to the labor income gives region's $i$ aggregate nominal income

$$Y_i = w_i L_i + R_i \tag{A.5}$$

*4. Equilibrium with transport costs*

As a result of (A.1) and the conditions of the demand for manufacturing varieties, the total demand faced by a firm $h$ located in region $i$ is

$$x_i(h) = \sum_{j \in \{B,H,W\}} (p_i(h))^{-\sigma} e_j q_j^{\sigma-1} T_{ij}^{1-\sigma} \tag{A.6}$$

where $T_{ij}$ is the trade cost from region $i$ to $j$, and $e_j$ is the aggregate level of expenditure on manufactures in region $j$. Then, from expressions (A.2), (A.3) and the normalizations in footnote (8), the profits of any manufacturing firm in region $i$ are

$$\pi_i = \frac{p_i}{\sigma}(x_i - 1) \tag{A.7}$$

The labor-market-clearing condition for any region $i$ results in the local labor supply being (from (A.2) and the profit-maximization conditions in the agricultural sector)

$$L_i = (1-\mu)\frac{C(x_i(h))}{w_i} n_i + K_i \left(\frac{\theta}{w_i}\right)^{\frac{1}{1-\theta}} \tag{A.8}$$

where the first (second) term of (A.8) is the manufacturing (agricultural) labor demand.

*5. Labor mobility*

Labor migration (and re-migration) is allowed in both directions between Border and West when a migration openness' indicator $m$ takes the value 1. Individuals have an idiosyncratic home-biased preference, represented by a parameter $\eta$, which implies they get a disutility from emigration. They do not emigrate unless such disutility is offset by real-wage differentials. For instance, a Border's inhabitant would migrate to the West if and only if $\frac{w_W}{q_W^\gamma} > \eta \frac{w_B}{q_B^\gamma}$, where $\frac{w_i}{q_i^\gamma}$ is region's $i$ real wage. As in Faini (1996), $\eta$ is distributed (across a block's population) according to a Pareto density function, $f(\eta) = \frac{\varepsilon}{\eta^{\varepsilon+1}}$ with $\eta \in (1, \infty)$ and $\varepsilon > 0$. Thus, the fraction of the original population from the Border that does not migrate is

$$\Psi = \min\left\{\int_{\frac{w_W}{q_W^\gamma}/\frac{w_B}{q_B^\gamma}}^{\infty} f(\eta) d\eta, 1\right\} = \min\left\{\left(\frac{\frac{w_B}{q_B^\gamma}}{\frac{w_W}{q_W^\gamma}}\right)^\varepsilon, 1\right\} \qquad (A.9)$$

and so, the percentage of the Border's population that migrates, $1 - \Psi$, increases with the wage differential. ε can be considered as an indicator of the degree of labor mobility. The migration West-Border is operationalized in the same way.

We have incorporated migration flows to our simulations, after calibrating one parameter (ε) measuring how big is the propensity to migrate of the Border population. Such calibration has followed Faini (1996)´s functional form.

Given that 1% of the Border population migrated to EU15 between 1990 and 2005, and prior to 1990 we can assume that Border-West migration was not allowed, in our model these facts imply that

$$0{,}01 = 1 - \Psi_{2005}$$

Let us denote by *r* to the ratio of Border-West GDP per capita in 2005. Then,

$$\varepsilon = \frac{\ln{(0{,}99)}}{\ln{(r)}} \approx 0{,}00672.$$